\documentclass[10pt,conference]{IEEEtran}
\IEEEoverridecommandlockouts
\usepackage{cite}
\usepackage{amsmath,amssymb,amsfonts}

\usepackage{graphicx}
\usepackage{textcomp}
\usepackage{xcolor}
\usepackage{multirow}

\usepackage{graphicx}
\usepackage{caption}
\usepackage{pgfplots}
\usepackage{tikz}
\usepackage{tkz-graph}
\usepackage{algorithmic}
\usepackage{algorithm}
\usepackage{hyperref}
\usepackage{cleveref}

\def\BibTeX{{\rm B\kern-.05em{\sc i\kern-.025em b}\kern-.08em
    T\kern-.1667em\lower.7ex\hbox{E}\kern-.125emX}}

\begin{document}

\title{Enhancing Routing in SD-EONs through Reinforcement Learning: A Comparative Analysis \\
\thanks{This paper was partially supported by NSF project award \#2008530.}}

\author{
    \IEEEauthorblockN{\hspace*{-2.9cm} Ryan McCann$^\dagger$}
    \IEEEauthorblockA{\hspace*{-2.9cm} \textit{ryan\_mccann@student.uml.edu}}
    \and
    \IEEEauthorblockN{Arash Rezaee$^\dagger$}
    \IEEEauthorblockA{\textit{arash\_rezaee@student.uml.edu}}
    \and
    \IEEEauthorblockN{Vinod M. Vokkarane$^\dagger$}
    \IEEEauthorblockA{\textit{vinod\_vokkarane@uml.edu}}
    \and
    \IEEEauthorblockA{\centering \textit{\vspace{-1.5em} $^\dagger$Electrical and Computer Engineering Department, University of Massachusetts Lowell, United States}}
}

\maketitle

\begin{abstract}
This paper presents an optimization framework for routing in software-defined elastic optical networks using reinforcement learning algorithms. We specifically implement and compare the epsilon-greedy bandit, upper confidence bound (UCB) bandit, and Q-learning algorithms to traditional methods such as K-Shortest Paths with First-Fit core and spectrum assignment (KSP-FF) and Shortest Path with First-Fit (SPF-FF) algorithms. Our results show that Q-learning significantly outperforms traditional methods, achieving a reduction in blocking probability (BP) of up to 58.8\% over KSP-FF, and 81.9\% over SPF-FF under lower traffic volumes. For higher traffic volumes, Q-learning maintains superior performance with BP reductions of 41.9\% over KSP-FF and 70.1\% over SPF-FF. These findings demonstrate the efficacy of reinforcement learning in enhancing network performance and resource utilization in dynamic and complex environments.

% Begin TODOs

% \textcolor{red}{Change all variables to italics.}

% \textcolor{red}{NSF project number.}

% \textcolor{red}{Q-Learning not as good in Erlang 500 as greedy bandit?}

% End TODOs

\end{abstract}


\section{Introduction}
The rapid advancement of technology and the exponential growth in data-intensive applications necessitate an in-depth understanding of the evolution and optimization of communication networks. The emergence of Elastic Optical Networks (EONs) and Software Defined Networking (SDN) represents a significant shift in this landscape, offering novel solutions to address the increasing demands for network flexibility and efficiency \cite{Gu2020MachineLearning, Cisco2020InternetReport}.

Optical networks, particularly those based on Wavelength Division Multiplexing (WDM), have traditionally played a pivotal role in global communication. However, the fixed channel spacings and limited adaptability of WDM systems are increasingly insufficient to meet the dynamic requirements of modern network traffic. In contrast, EONs, with their ability to provide dynamic bandwidth allocation and more granular channel spacings, offer a more efficient utilization of the optical spectrum and enhanced network performance \cite{Gerstel2012ElasticOpticalNetworking}.

The integration of Software Defined Networking in EONs (SD-EONs) has further advanced network management capabilities. SD-EONs' separation of control and data planes allows for a more flexible and efficient network configuration, which is crucial for managing the dynamic characteristics of EONs. This integration has led to improved network agility and an enhanced ability to respond to real-time changes in network conditions \cite{Nisar2020}.

Amidst these advancements, traditional routing and spectrum allocation methods in optical networks, such as the first-fit and shortest-path algorithms, have shown limitations in dealing with rapidly changing network scenarios due to their static nature and lack of adaptability to real-time conditions \cite{Chatter2018Frag}; to address these challenges, this paper explores three reinforcement learning algorithms—epsilon-greedy bandit, Upper Confidence Bound (UCB) bandit, and Q-learning—which span a spectrum of complexity from the straightforward epsilon-greedy bandit, balancing exploration and exploitation through random decisions based on past experiences, to the UCB bandit, using statistical confidence bounds to systematically explore and exploit network configurations, and finally to Q-learning, which incorporates considerations of future potential rewards into its iterative learning process, optimizing routing policies based on both immediate and long-term cumulative rewards in dynamic and stochastic environments \cite{watkins1989learning, Luong2019DeepRL}.

The primary aim of this study is to evaluate whether the increased complexity of the UCB bandit and Q-learning algorithms translates into enhanced performance in optimizing routing in SD-EONs compared to the epsilon-greedy bandit and traditional methods. By assessing their efficiency, adaptability, and scalability against each other and established baselines, this research seeks to provide insights into the trade-offs between algorithmic complexity and performance in modern network traffic scenarios.

The paper is structured as follows: 
Section \ref{sec:literature_review} reviews relevant literature on reinforcement learning in optical networking, 
Section \ref{sec:reinforcement_learning_algorithms} explains the background and functioning of the epsilon-greedy bandit, UCB bandit, and Q-learning algorithms, 
Section \ref{sec:proposed_algorithms} describes the proposed algorithms and their implementation for routing optimization, Section \ref{sec:system_architecture_assumptions} outlines the system architecture and assumptions,
and Section \ref{sec:results_discussion} presents the experimental results and comparative analysis. 
Finally, Section \ref{sec:conclusion} concludes the paper.

\section{Literature Review}
\label{sec:literature_review}

EONs, employing technologies like Orthogonal Frequency Division Multiplexing (OFDM) for its spectral efficiency and flexibility, and Flexgrid for future network architectures, dynamically allocate spectrum, advancing high-speed network design beyond traditional static management. However, their reliance on conventional network design principles, rather than artificial intelligence, highlights a growing need for more adaptive, AI-driven approaches to meet the increasing complexity and dynamic demands of modern EONs \cite{Chatter2015}.

Recent advancements in EONs have explored the use of Reinforcement Learning (RL), particularly Q-learning, for network optimization, marking a shift towards AI-based solutions. Unlike traditional network management techniques such as integer linear programming and heuristic approaches, RL offers efficient real-time responses to changing network conditions, paving the way for more adaptive and scalable network management solutions. The exploration of these networks extends to experimental demonstrations and assessing their practical viability, highlighting both achievements and challenges in elastic optical networking \cite{Luong2019DeepRL}.

Further advancements in SD-EONs have also explored the integration of Q-learning and hybrid approaches to improve Routing, Modulation, and Spectrum Assignment (RMSA). Bryant et al. discuss the application of off-policy Q-learning to enhance routing efficiency in optical networks, highlighting its effectiveness with a large number of k-paths \cite{Bryant2022QLearning}. Ríos-Villalba et al. present a hybrid approach combining Q-learning with k-shortest paths to address the RSA problem, demonstrating improved resource utilization and reduced latency \cite{RiosVillalba2023Hybrid}. Other studies explore deep reinforcement learning and multi-band networks, which are beyond the scope of this research \cite{Chen2019DeepRMSA, Sheikh2021MB, Errea2023DeepRL}.

Despite these advancements, several gaps remain. Bryant et al.'s study did not explore multiple reward policies or compare Q-learning with the traditional KSP algorithm. The computational feasibility of calculating numerous paths \textit{(10-20)} for every request remains questionable, and agent convergence and hyperparameter settings lack demonstration. Similarly, Ríos-Villalba et al. do not show conclusive results with a fully deployed Q-learning model or specify the number of paths considered.

No study has deployed a fully autonomous Q-learning model significantly outperforming traditional baselines with a moderate selection of paths, nor compared it to simpler bandit algorithms to justify Q-learning's complexity. Additionally, the impact of hyperparameter tuning and reward function shaping on reinforcement learning model performance remains underexplored.

\section{Reinforcement Learning Algorithms}
\label{sec:reinforcement_learning_algorithms}

\subsection{Introduction to Reinforcement Learning}

Reinforcement Learning (RL) is a type of machine learning where an agent learns to make decisions by interacting with an environment. The agent takes actions in the environment, receives feedback in the form of rewards, and updates its knowledge to improve future decision-making. The goal is to learn a policy that maximizes the cumulative reward over time. Key concepts in RL include the agent, which is the entity making decisions, and the environment, the system with which the agent interacts. The state (\(s\)) represents the current situation of the environment, and the action (\(a\)) is a decision made by the agent that affects the environment. The reward (\(r\)) is the feedback from the environment, indicating the immediate benefit of an action. The policy \(\pi(s, a)\) is a strategy used by the agent to determine actions based on the current state. The value function estimates the expected cumulative reward from a given state-action pair (Action Value Function \(Q(s, a)\)).

\subsection{Epsilon-Greedy Bandit}

The epsilon-greedy bandit algorithm is a simple RL method used in multi-armed bandit problems. It balances exploration (trying new actions to discover their rewards) and exploitation (choosing the best-known action to maximize reward). At each step, with probability \(\epsilon\), the agent selects a random action (exploration). With probability \(1 - \epsilon\), it selects the action with the highest estimated reward (exploitation). The value function in this context is the estimated reward for each action, typically maintained in a table \(Q(s, a)\). After taking action \(a\) in state \(s\) and receiving reward \(r\), the value estimate \(Q(s, a)\) is updated incrementally as shown by:

\begin{equation}
Q(s, a) \leftarrow Q(s, a) + \frac{1}{N(s, a)} (r - Q(s, a))
\label{eq:epsilon_greedy_update}
\end{equation}

where \(N(s, a)\) is the number of times action \(a\) has been taken in state \(s\). This update rule ensures that each new reward slightly adjusts the estimated value of the action, weighted by the inverse of the number of times the action has been selected, allowing the estimate to converge over time as more rewards are observed.

\subsection{Upper Confidence Bound (UCB) Bandit}

The UCB bandit algorithm is designed to address the exploration-exploitation trade-off by selecting actions based on both their estimated reward and the uncertainty (or confidence) in these estimates. At each step, the agent selects the action \(a\) that maximizes the following:

\begin{equation}
Q(s, a) + c \sqrt{\frac{\ln t}{N(s, a)}}
\label{eq:ucb_formula}
\end{equation}

where \(Q(s, a)\) is the estimated value for action \(a\), \(t\) is the total number of steps taken, \(s\) is the current state, \(N(s, a)\) is the number of times action \(a\) has been taken in state \(s\), and \(c\) is a parameter controlling the degree of exploration. Similar to the epsilon-greedy algorithm, \(Q(s, a)\) represents the estimated reward for each action in the current state. The value estimates are also updated by Equation \eqref{eq:epsilon_greedy_update} after action selection.

\subsection{Q-Learning}

Q-learning is a model-free RL algorithm that aims to learn the optimal policy by iteratively improving the estimates of the action-value function \(Q(s, a)\). At each step, the agent takes action \(a\) in state \(s\), observes the reward \(r\) and the next state \(s'\), and updates the action-value function \(Q(s, a)\) using the following rule:

\begin{equation}
Q(s, a) \leftarrow Q(s, a) + \alpha \left( r + \gamma \max_{a'} Q(s', a') - Q(s, a) \right)
\label{eq:q_learning_update}
\end{equation}

where \(\alpha\) is the learning rate and \(\gamma\) is the discount factor, representing the importance of max potential future rewards \(Q(s', a'\)), where \(s'\) is the next state and \(a'\) is the optimal action in \(s'\). \(Q(s, a)\) represents the expected cumulative reward of taking action \(a\) in state \(s\) and following the optimal policy thereafter. The policy \(\pi(s, a)\) is derived from the action-value function. Typically, the agent selects the action with the highest \(Q\)-value in each state. But, the Q-learning algorithm also depends on $\epsilon$ in order to act with some degree of randomness.

\section{Proposed Methodology}
\label{sec:proposed_algorithms}

This section outlines the methodology for optimizing SD-EON routing using the described RL algorithms.

\subsection{Epsilon-Greedy and UCB Bandit Algorithms}

Both the epsilon-greedy and UCB bandit algorithms are designed to optimize routing by selecting the best path from a set of pre-computed paths for each source-destination pair in the network topology. The pre-computed paths are generated using the KSP algorithm, providing \(k\) potential paths for each source-destination pair. The Q-table is constructed statically based on these pre-computed paths, but the paths are learned and selected dynamically by the reinforcement learning agent. After selecting a path, the agent sends it to the controller, which then uses the first-fit algorithm for core and spectrum assignment. If the request is blocked, a negative reward is returned to the agent. If the request is routed, a positive or non-negative reward is assigned.

Algorithm 1 outlines the pseudocode for the epsilon-greedy and UCB bandit algorithms as implemented in this research. The algorithm starts by initializing the Q-values for all state-action pairs and initializing the count of actions \(N(s, a)\) for all actions. The exploration rate \(\epsilon\) and the exploration parameter \(c\) are then set, the latter being specific to the UCB bandit.

Each episode begins with iterating over each request within the episode (Lines 1-2). For every request, a source-destination pair is chosen randomly and uniformly (Line 3). An action \(a\) is then chosen using the \(\epsilon\)-greedy policy: if a random number is less than \(\epsilon\) (Line 4), a random path \(a\) is selected (Line 5); otherwise, for the UCB algorithm, the path that maximizes the sum of the Q-value and the confidence bound is selected (Lines 7-8). If not using UCB, the path with the highest Q-value is selected (Lines 9-10).

The selected path \(a\) is then sent to the controller for assignment (Line 13). If the request is blocked, a reward \(R(s, a) = \textit{negative}\) is received (Lines 14-15); otherwise, a reward \(R(s, a) = \textit{non-negative}\) is received (Lines 16-17). The Q-value \(Q(s, a)\) is updated using the received reward (Line 19). Finally, the count \(N(s, a)\) for the action is incremented (Line 20).

\begin{algorithm}
\caption{Epsilon-Greedy and UCB Bandit Algorithms}
\begin{algorithmic}[1]
\REQUIRE Initialize \(Q(s, a)\) for all state-action pairs
\REQUIRE Initialize \(N(s, a)\) for all actions
\REQUIRE Set exploration rate \(\epsilon\) and exploration parameter \(c\)
\FOR{each episode}
    \FOR{each request}
        \STATE Choose a source-destination pair randomly and uniformly (state \(s\))
        \IF{random number \( < \epsilon\)} 
            \STATE Select a random path \(a\)
        \ELSE
            \IF{UCB}
                \STATE Select \(a = \arg\max \left( Q(s, a) + c \sqrt{\frac{\ln t}{N(s, a)}} \right)\)
            \ELSE
                \STATE Select \(a = \arg\max Q(s, a)\)
            \ENDIF
        \ENDIF
        \STATE Send path \(a\) to controller for assignment
        \IF{request is blocked}
            \STATE Receive reward \(R(s, a) = \textit{negative}\)
        \ELSE
            \STATE Receive reward \(R(s, a) = \textit{non-negative}\)
        \ENDIF
        \STATE Update \(Q(s, a) \leftarrow Q(s, a) + \frac{1}{N(s, a)} (r - Q(s, a))\)
        \STATE Increment \(N(s, a)\)
    \ENDFOR
\ENDFOR
\end{algorithmic}
\end{algorithm}

The time complexity for the overall process includes components done once and those done for each request. Path computation using the k-shortest path algorithm for each source-destination pair, done once, is \(O(S \cdot k \cdot (E + V \cdot \log V))\), where \(S\) is the number of source-destination pairs, \(k\) is the number of paths, \(V\) is the number of vertices, and \(E\) is the number of edges. Q-table initialization, also done once, has a complexity of \(O(S \cdot k)\). For every request, the process includes looking up the paths for the source and destination with a complexity of \(O(1)\) and evaluating and selecting a path using the epsilon-greedy or UCB bandit algorithms with a complexity of \(O(k)\). The UCB bandit algorithm is slightly more memory intensive due to the additional calculation of the confidence bound.

\subsection{Q-Learning Algorithm}

The Q-learning algorithm is designed to optimize routing by learning from the congestion levels in the network. For each source and destination node pair in the network topology, there are \(k\) pre-computed paths the agent may select from, a very similar setup to the previously mentioned bandit algorithms. Unlike the simpler bandit algorithms, Q-learning also takes into account the congestion of a path before and after allocation, allowing the agent to understand and adapt to network conditions over time.

The state \(s\) is defined based on the congestion level of the network before the allocation of a new request. Congestion is calculated as the average congestion along all the cores and links of a path, defined as the number of spectral slots occupied divided by the total number of free spectral slots. The congestion state is categorized into two levels: Level 1 for congestion below 0.3 (30\%) and Level 2 for congestion above or equal to 0.3 (30\%). Each congestion level accesses the same set of \(k\) pre-computed paths, but the Q-values associated with these paths differ for each congestion level. This is because a path's performance can vary depending on the congestion level, making it better or worse at different levels of congestion.

The action \(a\) corresponds to selecting one of the \(k\) pre-computed paths for a given source-destination pair. After selecting a path and observing the result (routed or blocked), the congestion level is re-measured. This post-allocation congestion state is used to update the Q-value. Specifically, the Q-value \(Q(s, a)\) is updated using the maximum future Q-value from the new state \(s'\), which reflects the congestion level after allocation. The update rule for Q-learning is later used, which is shown in Equation \eqref{eq:q_learning_update}.

Algorithm 2 outlines the pseudocode for the Q-learning process as implemented in this paper. Each episode begins with iterating over each request within the episode (Lines 1-2). For every request, the initial state \(s\) (pre-allocation congestion level for the current source-destination pair) is observed (Line 3). The state includes the current source and destination pair for the request, as well as the congestion levels for each of the \(k\) pre-computed paths with respect to their current congestion. An action \(a\) is chosen from these paths using the \(\epsilon\)-greedy policy based on their Q-values in the current congestive state (Lines 4-8). The chosen action \(a\) is executed by sending the request to the controller (Line 9). If the request is routed, a reward \(R(s, a) = \textit{non-negative}\) is received (Lines 10-11); otherwise, a reward \(R(s, a) = \textit{negative}\) is received (Lines 12-13). The new state \(s'\) (post-allocation congestion level for the selected path) is observed (Line 15). The Q-value \(Q(s, a)\) is updated using Equation \eqref{eq:q_learning_update} (Lines 16-17). This process continues for each request in the current episode until all requests are processed.

\begin{algorithm}[h]
\caption{Q-Learning Algorithm}
\begin{algorithmic}[1] % The [1] ensures line numbers are included
\REQUIRE Initialize \(Q(s, a)\) to zero for all state-action pairs
\REQUIRE Set learn rate \(\alpha\), disc. factor \(\gamma\), and exploration \(\epsilon\)
\FOR{each episode}
    \FOR{each request}
        \STATE Observe initial state \(s\) (pre-allocation congestion level for current source-destination pair)
        \IF{ random number \( < \epsilon\)}
            \STATE Choose action \(a\) randomly
        \ELSE
            \STATE Choose action \(a = \arg\max Q(s, a)\)
        \ENDIF
        \STATE Execute action \(a\) (send path to controller)
        \IF{request is routed}
            \STATE Receive reward \(R(s, a) = \textit{non-negative}\)
        \ELSE
            \STATE Receive reward \(R(s, a) = \textit{negative}\)
        \ENDIF
        \STATE Observe new state \(s'\) (post-allocation congestion level for the selected path)
        \STATE Update \(Q(s, a)\) using:
        \STATE \(\begin{aligned}[t]
        Q(s, a) \leftarrow & Q(s, a) + \alpha \big( R(s, a) + \gamma \max_{a'} Q(s', a') \\
        & - Q(s, a) \big)
        \end{aligned}\)
    \ENDFOR
\ENDFOR
\end{algorithmic}
\end{algorithm}

The complexity of the Q-learning algorithm involves first considering the source-destination pair for the current request. For each source-destination pair, the algorithm considers the \(k\) pre-computed paths. The congestion state, which has two possible levels, is computed based on the current congestion along these paths. The action involves selecting one of these paths based on their Q-values in the current state. Thus, the overall time complexities are: path computation \(O(S \cdot k \cdot (E + V \cdot \log(V)))\) and Q-table initialization \(O(S \cdot k \cdot C)\) done once, and source-destination lookup \(O(1)\) and path evaluation and selection \(O(C \times k)\) for Q-learning for each request, where $C$ is the number of congestion levels.

\section{System Architecture and Assumptions}
\label{sec:system_architecture_assumptions}

\begin{figure}[h!]
    \centering
    % nsfnet.tex

\begin{tikzpicture}[scale=0.8]
\tikzstyle{vertex}=[circle, draw, fill=cyan, text=black, minimum size=13pt, inner sep=0pt, font=\small]
    \tikzstyle{edge}=[draw, thick]
    \tikzstyle{edgelabel}=[fill=white, inner sep=1pt, font=\footnotesize]

    \node[vertex] (1) at (-10, 3) {1};
    \node[vertex] (2) at (-11, 0.4) {2};
    \node[vertex] (3) at (-8, -0.5) {3};
    \node[vertex] (4) at (-8.2, 1.7) {4};
    \node[vertex] (5) at (-6.8, 1.2) {5};
    \node[vertex] (6) at (-5.7, -1) {6};
    \node[vertex] (7) at (-5.5, 0.5) {7};
    \node[vertex] (8) at (-5.0, 1.85) {8};
    \node[vertex] (9) at (-3.5, -1.0) {9};
    \node[vertex] (10) at (-3.4, 1.65) {10};
    \node[vertex] (11) at (-4, 2.8) {11};
    \node[vertex] (12) at (-2.6, 0.3) {12};
    \node[vertex] (13) at (-1.8, 2.5) {13};
    \node[vertex] (14) at (-1.5, 1.7) {14};

    % Define edges
    \path[edge] (1) -- (2) node[edgelabel, midway, above] {1000};
    \path[edge] (3) -- (2) node[edgelabel, midway, below] {600};
    \path[edge] (3) -- (1) node[edgelabel, midway, above left] {1500};
    \path[edge] (2) -- (4) node[edgelabel, midway, below] {700};
    \path[edge] (5) -- (4) node[edgelabel, midway, below] {600};
    \path[edge] (3) -- (6) node[edgelabel, midway, above] {1800};
    \path[edge] (5) -- (6) node[edgelabel, midway, above] {1200};
    \path[edge] (5) -- (7) node[edgelabel, midway, above] {600};
    \path[edge] (8) -- (7) node[edgelabel, midway, below] {700};
    \path[edge] (9) -- (7) node[edgelabel, midway, above] {1300};
    \path[edge] (9) -- (6) node[edgelabel, midway, below] {1000};
    \path[edge] (12) -- (6) node[edgelabel, midway, below] {1800};
    \path[edge] (8) -- (10) node[edgelabel, midway, below] {700};
    \path[edge] (9) -- (10) node[edgelabel, midway, above] {700};
    \path[edge] (13) -- (10) node[edgelabel, midway, below] {300};
    \path[edge] (14) -- (10) node[edgelabel, midway, below left] {300};
    \path[edge] (14) -- (11) node[edgelabel, midway, above] {700};
    \path[edge] (13) -- (11) node[edgelabel, midway, above] {600};
    \path[edge] (13) -- (12) node[edgelabel, midway, below] {300};
    \path[edge] (14) -- (12) node[edgelabel, midway, below] {100};
    \path[edge] (11) -- (4) node[edgelabel, midway, above] {1900};
    \path[edge] (8) -- (1) node[edgelabel, midway, above] {2400};
\end{tikzpicture}
    \caption{NSFNet Topology.}
    \label{fig:nsfnet}
\end{figure}
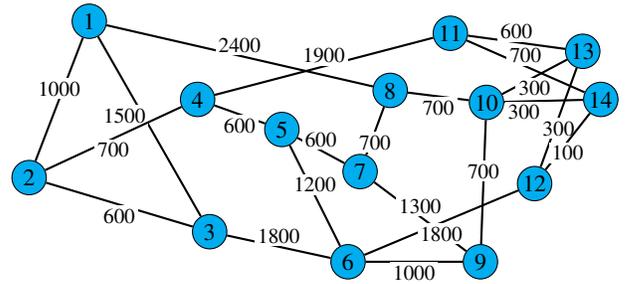

The foundation of this research is a dynamic simulation model of a SD-EON, constructed in Python \cite{SDN_Simulator}. Based on concepts from a doctoral dissertation \cite{wang2022dynamic}, this model utilizes the NSFNet topology with 14 nodes and 22 bi-directional links, as depicted in Figure \ref{fig:nsfnet}.

Each link in the simulation comprises of four cores, with 128 spectral slots per core. Network conditions are simulated using requests with bit rates of 25, 50, and 100 Gbps, distributed in a 3:5:2 ratio. While each lightpath requires a guard band, these are not included in Table \ref{tab:modulation_formats}, which details specific bit rates, spectral slot requirements, and supported distances for each modulation format. It should be noted that we adopt bandwidth-tunable transceivers with a slot spectral width of 12.5 GHz.

\begin{table}
\centering
% modulation_formats.tex

\begin{tabular}{|c|c|c|c|}
\hline
\textbf{Mod Format} & \textbf{Bit Rate (Gbps)} & \textbf{Slots} & \textbf{Distance (KM)} \\
\hline
\multirow{3}{*}{QPSK}   & 25  & 1 & 22,160 \\
                        & 50  & 2 & 11,080 \\
                        & 100 & 4 & 5,540 \\
\hline
\multirow{3}{*}{16-QAM} & 25  & 1 & 9,500 \\
                        & 50  & 1 & 4,750 \\
                        & 100 & 2 & 2,375 \\
\hline
\multirow{3}{*}{64-QAM} & 25  & 1 & 3,664 \\
                        & 50  & 1 & 1,832 \\
                        & 100 & 2 & 916 \\
\hline
\end{tabular}

\caption{Modulation Formats with Bit Rates and Slots}
\label{tab:modulation_formats}
\end{table}

Request arrival and holding times follow exponential and Poisson random distributions, respectively, with source-destination pairs determined uniformly. The arrival rate is normalized by the number of cores per link, and each simulation episode is defined as 2,000 requests. A constant mean holding time of 5 time units is used for each request generated.

\section{Results and Discussion}
\label{sec:results_discussion}

This section presents the experimental results of the proposed reinforcement learning algorithms—epsilon-greedy bandit, UCB bandit, and Q-learning—compared to traditional K-Shortest Paths with First-Fit (KSP-FF) $k=3$, Shortest Path with First-Fit (SPF-FF), and KSP-FF with $k=inf$ (KSP-$inf$) algorithms. The experiments were conducted using Erlang values of 500, 750, and 1000 to evaluate the reduction in blocking probability (BP) achieved by the proposed methods.

The KSP-FF and SPF-FF algorithms select paths based on their link lengths, favoring shorter paths to minimize resource utilization. Regarding time complexity, the KSP-FF algorithm has a time complexity of \(O(k \cdot (E + V \cdot \log V))\), where \(E\) is the number of edges and \(V\) is the number of vertices in the network. The SPF-FF algorithm has a time complexity of \(O(E + V \cdot \log V)\), as it involves a single shortest path computation.

\subsection{Results}

For Erlang 500, the RL algorithms exhibited improvements in reducing BP over the baselines, shown in Fig. \ref{fig:e500_nsfnet_k3}. The epsilon-greedy bandit algorithm, configured with an epsilon of 1\%, a positive reward of 1, and a negative reward of 100, led to a reduction in BP of 98\% over KSP-FF, 99\% over SPF-FF, and 84.6\% over KSP-$inf$. The UCB bandit algorithm, configured with an epsilon of 20\%, a confidence interval value of 2, a positive reward of 1, and a negative reward of 10, resulted in a reduction in BP of 30\% over KSP-FF and 73\% over SPF-FF, but showed an increase in BP of 81\% over KSP-$inf$. The Q-learning algorithm, with an epsilon linearly decaying from 10\% to 5\%, a positive reward of 1, and a negative reward of 100, showed a reduction in BP of 92\% over KSP-FF, 97\% over SPF-FF, and 38.5\% over KSP-$inf$. The learning rate was set to 0.05, and the discount factor was 0.01.

The results for Erlang 750, as displayed in Fig. \ref{fig:e750_nsfnet_k3}, demonstrate improvements by the RL algorithms. The epsilon-greedy bandit algorithm, with an epsilon of 6\%, a positive reward of 10, and a negative reward of 100, led to a reduction in BP of 36.9\% over KSP-FF, 72.3\% over SPF-FF, and an increase of 28.2\% over KSP-$inf$. The UCB bandit algorithm, configured with an epsilon of 10\%, a confidence interval value of 2, a positive reward of 10, and a negative reward of 10, resulted in a reduction in BP of 35.4\% over KSP-FF, 71.6\% over SPF-FF, and an increase of 25.6\% over KSP-$inf$. The Q-learning algorithm, with an epsilon linearly decaying from 20\% to 5\%, a positive reward of 10, and a negative reward of 100, showed a reduction in BP of 58.8\% over KSP-FF, 81.9\% over SPF-FF, and 15\% over KSP-$inf$. The learning rate was 0.01, and the discount factor was 0.95.

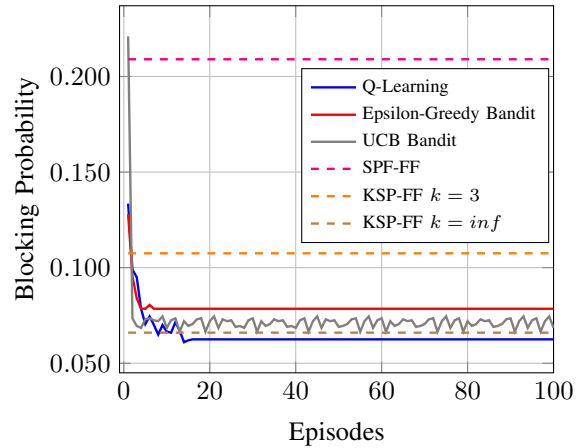
\begin{figure}[h!]
    \centering
    % k3.tex

\begin{tikzpicture}
\begin{axis}[
    title={Average Blocking Probability vs. Episodes},
    xlabel={Episodes},
    ylabel={Blocking Probability},
    grid,
    scaled y ticks=false,
    yticklabel style={
        /pgf/number format/fixed,
        /pgf/number format/fixed zerofill,
        /pgf/number format/precision=3,
    },
    xmax=100,
    xmin=-1,
    width=0.83\linewidth,
    height=0.73\linewidth,
    legend style={
        at={(0.99,0.83)},
        anchor=north east,
        legend cell align=left,
        font=\scriptsize,
    },
]

\addplot[
    no markers,
    solid,
    color=blue,
    line width=0.9pt,
] table [x=episodes, y=q_learning, col sep=comma] {data/erlang_1000/k3/NSFNet/best_performing.csv};
\addlegendentry{Q-Learning}

\addplot[
    no markers,
    solid,
    color=red,
    line width=0.9pt,
] table [x=episodes, y=epsilon_greedy_bandit, col sep=comma] {data/erlang_1000/k3/NSFNet/best_performing.csv};
\addlegendentry{Epsilon-Greedy Bandit}

\addplot[
    no markers,
    solid,
    color=gray,
    line width=0.9pt,
] table [x=episodes, y=ucb_bandit, col sep=comma] {data/erlang_1000/k3/NSFNet/best_performing.csv};
\addlegendentry{UCB Bandit}

\addplot[
    no markers,
    dashed,
    color=magenta,
    line width=0.9pt,
] table [x=episodes, y=spf, col sep=comma] {data/erlang_1000/k3/NSFNet/best_performing.csv};
\addlegendentry{SPF-FF}

\addplot[
    no markers,
    dashed,
    color=orange,
    line width=0.9pt,
] table [x=episodes, y=ksp, col sep=comma] {data/erlang_1000/k3/NSFNet/best_performing.csv};
\addlegendentry{KSP-FF $k=3$}

\addplot[
    no markers,
    dashed,
    color=brown,
    line width=0.9pt,
] table [x=episodes, y=ffll, col sep=comma] {data/erlang_1000/k3/NSFNet/best_performing_v2.csv};
\addlegendentry{KSP-FF $k=inf$}

\end{axis}
\end{tikzpicture}
    \caption{BP vs. Episodes $Erlang=1000$}
    \label{fig:e1000_nsfnet_k3}
\end{figure}

For Erlang 1000, the RL algorithms also outperformed the traditional baselines, though the degree of improvement varied, as shown in Fig. \ref{fig:e1000_nsfnet_k3}. The epsilon-greedy bandit algorithm, configured with an epsilon of 1\%, a non-negative reward of 0, and a negative reward of 10, showed a reduction in BP of 27\% over KSP-FF, 62.4\% over SPF-FF, and an increase of 19.7\% over KSP-$inf$. The UCB bandit algorithm, with an epsilon of 20\%, a positive reward of 10, and a negative reward of 10, led to a reduction in BP of 34\% over KSP-FF, 66.1\% over SPF-FF, and an increase of 4.5\% over KSP-$inf$. The Q-learning algorithm, with a constant epsilon of 5\%, a positive reward of 1, and a negative reward of 10, resulted in a reduction in BP of 41.9\% over KSP-FF, 70.1\% over SPF-FF, and 5.3\% over KSP-$inf$. The learning rate was set to 0.05, and the discount factor was 0.01.

\subsection{Discussion}

The results demonstrate that reinforcement learning algorithms, particularly Q-learning at higher traffic volumes and the Epsilon-greedy bandit at lower ones, significantly reduce BP in SD-EONs compared to traditional methods. For Erlang 500, as shown in Fig. \ref{fig:e500_nsfnet_k3}, Q-learning's superior performance is due to its ability to adapt to network conditions over time, considering both immediate and future rewards, with a decaying $\epsilon$ strategy enhancing exploration. The epsilon-greedy bandit algorithm showed the most significant BP reductions, likely due to its simplicity being optimal at low traffic volumes, though Q-learning came very close.

Typically, the bandit algorithms converge faster than Q-learning, but over time, Q-learning usually converges to a lower blocking probability. However, Q-learning does not reach a lower blocking probability compared to the Epsilon-greedy bandit at a lower traffic volume, signifying that as the problem becomes simpler (with fewer requests), a less complex bandit algorithm proves its importance.

At Erlang 750, Q-learning began to outperform every algorithm, benefiting from its dynamic learning approach, reducing BP by 15\% over KSP-$inf$. For Erlang 1000, Q-learning remained robust, maintaining significant BP reductions, outperforming KSP-$inf$ by 5.3\%. The UCB bandit algorithm, while less effective than Q-learning, showed consistent improvements due to systematic exploration, and the epsilon-greedy algorithm also achieved considerable BP reductions.

Hyperparameters were tuned with 150-200 configurations for each algorithm at each traffic volume, ensuring optimal performance across a total of 450-600 configurations. Reward shaping was also crucial for effective learning. Despite the increased complexity of Q-learning, its significant BP reduction justifies its use in dynamic and complex network environments with higher traffic.

% Begin NSFNet Figures

\begin{figure}[h!]
    \centering
    % k3.tex

\begin{tikzpicture}
\begin{axis}[
    title={Average Blocking Probability vs. Episodes},
    xlabel={Episodes},
    ylabel={Blocking Probability},
    grid,
    scaled y ticks=false,
    yticklabel style={
        /pgf/number format/fixed,
        /pgf/number format/fixed zerofill,
        /pgf/number format/precision=3,
    },
    xmax=100,
    xmin=-1,
    ymax=0.135,
    ymin=-0.01,
    width=0.83\linewidth,
    height=0.73\linewidth,
    legend style={
        at={(0.99,0.93)},
        anchor=north east,
        legend cell align=left,
        font=\scriptsize,
    },
]

\addplot[
    no markers,
    solid,
    color=blue,
    line width=0.9pt,
] table [x=episodes, y=q_learning, col sep=comma] {data/erlang_500/k3/NSFNet/best_performing.csv};
\addlegendentry{Q-Learning}

\addplot[
    no markers,
    solid,
    color=red,
    line width=0.9pt,
] table [x=episodes, y=epsilon_greedy_bandit, col sep=comma] {data/erlang_500/k3/NSFNet/best_performing.csv};
\addlegendentry{Epsilon-Greedy Bandit}

\addplot[
    no markers,
    solid,
    color=gray,
    line width=0.9pt,
] table [x=episodes, y=ucb_bandit, col sep=comma] {data/erlang_500/k3/NSFNet/best_performing.csv};
\addlegendentry{UCB Bandit}

\addplot[
    no markers,
    dashed,
    color=magenta,
    line width=0.9pt,
] table [x=episodes, y=spf, col sep=comma] {data/erlang_500/k3/NSFNet/best_performing.csv};
\addlegendentry{SPF-FF}

\addplot[
    no markers,
    dashed,
    color=orange,
    line width=0.9pt,
] table [x=episodes, y=ksp, col sep=comma] {data/erlang_500/k3/NSFNet/best_performing.csv};
\addlegendentry{KSP-FF $k=3$}

\addplot[
    no markers,
    dashed,
    color=brown,
    line width=0.9pt,
] table [x=episodes, y=ffll, col sep=comma] {data/erlang_500/k3/NSFNet/best_performing.csv};
\addlegendentry{KSP-FF $k=inf$}

\end{axis}
\end{tikzpicture}
    \caption{BP vs. Episodes $Erlang=500$}
    \label{fig:e500_nsfnet_k3}
\end{figure}
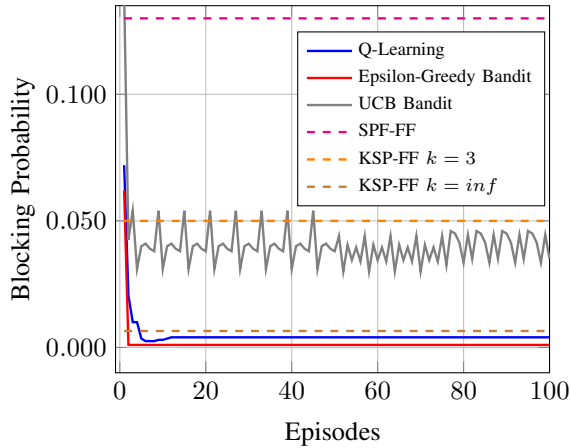

\begin{figure}[h!]
    \centering
    % k3.tex

\begin{tikzpicture}
\begin{axis}[
    title={Average Blocking Probability vs. Episodes},
    xlabel={Episodes},
    ylabel={Blocking Probability},
    grid,
    scaled y ticks=false,
    yticklabel style={
        /pgf/number format/fixed,
        /pgf/number format/fixed zerofill,
        /pgf/number format/precision=3,
    },
    xmax=100,
    xmin=-1,
    width=0.83\linewidth,
    height=0.73\linewidth,
    legend style={
        at={(0.99,0.85)},
        anchor=north east,
        legend cell align=left,
        font=\scriptsize,
    },
]

\addplot[
    no markers,
    solid,
    color=blue,
    line width=0.9pt,
] table [x=episodes, y=q_learning, col sep=comma] {data/erlang_750/k3/NSFNet/best_performing.csv};
\addlegendentry{Q-Learning}

\addplot[
    no markers,
    solid,
    color=red,
    line width=0.9pt,
] table [x=episodes, y=epsilon_greedy_bandit, col sep=comma] {data/erlang_750/k3/NSFNet/best_performing.csv};
\addlegendentry{Epsilon-Greedy Bandit}

\addplot[
    no markers,
    solid,
    color=gray,
    line width=0.9pt,
] table [x=episodes, y=ucb_bandit, col sep=comma] {data/erlang_750/k3/NSFNet/best_performing.csv};
\addlegendentry{UCB Bandit}

\addplot[
    no markers,
    dashed,
    color=magenta,
    line width=0.9pt,
] table [x=episodes, y=spf, col sep=comma] {data/erlang_750/k3/NSFNet/best_performing.csv};
\addlegendentry{SPF-FF}

\addplot[
    no markers,
    dashed,
    color=orange,
    line width=0.9pt,
] table [x=episodes, y=ksp, col sep=comma] {data/erlang_750/k3/NSFNet/best_performing.csv};
\addlegendentry{KSP-FF $k=3$}

\addplot[
    no markers,
    dashed,
    color=brown,
    line width=0.9pt,
] table [x=episodes, y=ffll, col sep=comma] {data/erlang_750/k3/NSFNet/best_performing_v2.csv};
\addlegendentry{KSP-FF $k=inf$}

\end{axis}
\end{tikzpicture}
    \caption{BP vs. Episodes $Erlang=750$}
    \label{fig:e750_nsfnet_k3}
\end{figure}

% End NSFNet Figures

\section{Conclusion}
\label{sec:conclusion}

% In this study, we explored the optimization of routing in SD-EONs using reinforcement learning algorithms. Among the methods investigated, Q-learning emerged as the best-performing algorithm as Erlang values increased, significantly outperforming simpler bandit algorithms such as epsilon-greedy and UCB bandits. Q-learning's ability to consider both immediate and future rewards enabled it to adapt more effectively to dynamic network conditions, resulting in superior performance in reducing blocking probability. Notably, Q-learning performed better than KSP-$inf$, despite KSP-$inf$ potentially considering every possible path. Future work will extend this research to include core and spectrum assignment, further enhancing network optimization. Additionally, we plan to implement deep learning techniques to potentially improve the efficiency and scalability of RL approaches in SD-EONs.

% Remember, this was changed for arxiv!

In this study, we explored the optimization of routing in SD-EONs using reinforcement learning algorithms. Q-learning emerged as the best-performing algorithm as Erlang values increased, significantly outperforming simpler bandit algorithms such as epsilon-greedy and UCB bandits. Its ability to consider both immediate and future rewards enabled it to adapt more effectively to dynamic network conditions, resulting in superior performance in reducing blocking probability. Future work will extend this research to include core and spectrum assignment and explore deep learning techniques to improve the efficiency and scalability of RL approaches in SD-EONs.

\bibliographystyle{IEEEtran}
\bibliography{references.bib}
\vspace{12pt}

\end{document}